\documentclass[12pt]{article}
\usepackage{amssymb}

\usepackage[T2A]{fontenc} 
\usepackage[cp1251]{inputenc}
\usepackage[english]{babel}

\def\hybrid{\topmargin 0pt      \oddsidemargin 0pt
        \headheight 0pt \headsep 0pt
        \voffset=-0.5cm
        \hoffset=-0.25in
        \textwidth 6.75in
        \textheight 9.5in       
        \marginparwidth 0.0in
        \parskip 5pt plus 1pt   \jot = 1.5ex}
\catcode`\@=11
\def\marginnote#1{}

\newcount\hour
\newcount\minute
\newtoks\amorpm
\hour=\time\divide\hour by60 \minute=\time{\multiply\hour by60
\global\advance\minute by-\hour}
\edef\standardtime{{\ifnum\hour<12 \global\amorpm={am}%
        \else\global\amorpm={pm}\advance\hour by-12 \fi
        \ifnum\hour=0 \hour=12 \fi
        \number\hour:\ifnum\minute<10 0\fi\number\minute\the\amorpm}}
\edef\militarytime{\number\hour:\ifnum\minute<10 0\fi\number\minute}

\def\draftlabel#1{{\@bsphack\if@filesw {\let\thepage\relax
   \xdef\@gtempa{\write\@auxout{\string
      \newlabel{#1}{{\@currentlabel}{\thepage}}}}}\@gtempa
   \if@nobreak \ifvmode\nobreak\fi\fi\fi\@esphack}
        \gdef\@eqnlabel{#1}}
\def\@eqnlabel{}
\def\@vacuum{}
\def\draftmarginnote#1{\marginpar{\raggedright\scriptsize\tt#1}}
\def\draftlabel#1{{\@bsphack\if@filesw {\let\thepage\relax
   \xdef\@gtempa{\write\@auxout{\string
      \newlabel{#1}{{\@currentlabel}{\thepage}}}}}\@gtempa
   \if@nobreak \ifvmode\nobreak\fi\fi\fi\@esphack}
        \gdef\@eqnlabel{#1}}
\def\@eqnlabel{}
\def\@vacuum{}
\def\draftmarginnote#1{\marginpar{\raggedright\scriptsize\tt#1}}

\def\draft{\oddsidemargin -.5truein
        \def\@oddfoot{\sl preliminary draft \hfil
        \rm\thepage\hfil\sl\today\quad\militarytime}
        \let\@evenfoot\@oddfoot \overfullrule 3pt
        \let\label=\draftlabel
        \let\marginnote=\draftmarginnote
   \def\@eqnnum{(\theequation)\rlap{\kern\marginparsep\tt\@eqnlabel}%
\global\let\@eqnlabel\@vacuum}  }

\def\numberbysection{\@addtoreset{equation}{section}
        \def\theequation{\thesection.\arabic{equation}}}

\def\underline#1{\relax\ifmmode\@@underline#1\else
        $\@@underline{\hbox{#1}}$\relax\fi}

\def\titlepage{\@restonecolfalse\if@twocolumn\@restonecoltrue\onecolumn
     \else \newpage \fi \thispagestyle{empty}\c@page\z@
        \def\thefootnote{\fnsymbol{footnote}} }

\def\endtitlepage{\if@restonecol\twocolumn \else  \fi
        \def\thefootnote{\arabic{footnote}}
        \setcounter{footnote}{0}}  

\numberbysection \hybrid

\newcounter{mo}

\newcommand{\tr}{{\rm tr}}

\newcommand{\mL}{{\mathcal L}}
\newcommand{\mM}{{\mathcal M}}

\newcommand{\vth}{\vartheta}

\newcommand{\Mat}{ {\rm Mat}(N,\mathbb C) }
\newcommand{\MatM}{ {\rm Mat}(M,\mathbb C) }
\newcommand{\MatNM}{ {\rm Mat}(NM,\mathbb C) }

\newcommand{\mS}{\mathcal S}

\def\beq{\begin{equation}}
\def\eq{\end{equation}}
\def\p{\partial}

\def\res{\mathop{\hbox{Res}}\limits}

\begin{document}

\setcounter{page}{1}

\date{}
\date{}
\vspace{50mm}

\begin{flushright}
\end{flushright}
\vspace{0mm}

\begin{center}
\vspace{0mm} {\LARGE{Integrable System of Generalized }}
 \\ \vspace{4mm}
 {\LARGE{Relativistic Interacting Tops}}
\\
\vspace{14mm} {\large {I. Sechin}}{\small $^{\diamondsuit,\flat }$}\ \ \ \ \ \
 {\large {A. Zotov}}{\small $^{\diamondsuit}$}\\

 \vspace{6mm}
$^\diamondsuit$ -- {{\rm Steklov Mathematical Institute of Russian Academy of Sciences, \\
Gubkina str. 8, Moscow, 119991, Russia.}}
\vspace{1mm}

$^\flat$ -- {\rm Center for Advanced Studies,
Skolkovo Institute of Science and Technology, \\ Nobel str. 1, Moscow, 143026, Russia.
}
\end{center}
%
\begin{center}\small{{\rm E-mails:}{\rm\ \ shnbuz@gmail.com\,,\
 zotov@mi-ras.ru}}\end{center}

 \begin{abstract}
A family of integrable $GL(NM)$ models is described. On the one hand it generalizes the classical spin Ruijsenaars--Schneider systems (the case $N=1$), and on the other hand it generalizes the relativistic integrable tops on $GL(N)$ Lie group (the case $M=1$). The described models are obtained by means of the Lax pair with spectral parameter. Equations of motion are derived. For the construction of the Lax representation the $GL(N)$ $R$--matrix in the fundamental representation of $GL(N)$ is used.
 \end{abstract}



\bigskip


\section{Introduction}\label{sect1}
\setcounter{equation}{0}
This paper is a continuation of series of articles
 \cite{GSZ,SeZ,Z19}, where known integrable systems and related structures are extended through the use of quantum $R$--matrices
 (in the fundamental representation of $GL(N)$ Lie groups) being interpreted as matrix generalizations of the Kronecker function. It is convenient to give its explicit form the very beginning in the rational, trigonometric and elliptic cases since the identities to be used are valid in all these cases:

 \beq\label{a001}
  \begin{array}{c}
  \displaystyle{
 \phi(z,q)=\left\{
   \begin{array}{l}
    1/z+1/q\,,
    \\ \ \\
    \coth(z)+\coth(q)\,,
    \\ \ \\
      \displaystyle{
    \frac{\vth'(0)\vth(z+q)}{\vth(z)\vth(q)}\,.}
   \end{array}
 \right.
 \
 E_1(z)=\left\{
   \begin{array}{l}
 1/z\,,
\\ \ \\
   \coth(z)\,,
\\ \ \\
      \displaystyle{    \frac{\vth'(z)}{\vth(z)} }
   \end{array}
 \right.
 \
 \wp(z)=\left\{
   \begin{array}{l}
 1/z^2\,,
\\ \ \\
   1/\sinh^2(z)+\frac{1}{3}\,,
\\ \ \\
      \displaystyle{    -E_1'(z)+\frac{1}{3}\frac{\vth'''(0)}{\vth'(0)} }
   \end{array}
 \right.
 }
 \end{array}
 \eq
 All the functions are complex valued. So that the trigonometric and hyperbolic cases are actually the same.
In~(\ref{a001})  for all three cases we also give definitions of the first Eisenstein function $E_1(z)$ and the Weierstrass $\wp$-function. They appear in the expansion of $\phi(z,q)$ near its simple pole (with residue equal to one) at~$z=0$:
 \beq\label{a002}
  \begin{array}{c}
  \displaystyle{
 \phi(z,q)=z^{-1}+E_1(q)+z\,
 (E_1^2(q)-\wp(q))/2+O(z^2)\,.
 }
 \end{array}
 \eq
  The definitions and properties of elliptic functions can be found in \cite{Weil} (see also Appendix from \cite{Z19}).

For each of three cases the Kronecker function satisfies the summation formula --- the Fay identity of genus $1$:
  \beq\label{a003}
  \begin{array}{c}
  \displaystyle{
\phi(z_1,q_1)\phi(z_2,q_2)=\phi(z_1-z_2,q_1)\phi(z_2,q_1+q_2)+\phi(z_2-z_1,q_2)\phi(z_1,q_1+q_2)\,,
 }
 \end{array}
 \eq
as well as its degenerations corresponding to equal arguments:
   \beq\label{a004}
  \begin{array}{c}
  \displaystyle{
 \phi(z,q_1)\phi(z,q_2)=\phi(z,q_1+q_2)(E_1(z)+E_1(q_1)+E_1(q_2)-E_1(q_1+q_2+z))\,,
 }
 \end{array}
 \eq
   \beq\label{a005}
  \begin{array}{c}
  \displaystyle{
 \phi(z,q)\phi(z,-q)=\wp(z)-\wp(q)\,.
 }
 \end{array}
 \eq

The Fay identity (\ref{a003}) can be considered as a particular scalar case of the associative Yang--Baxter equation \cite{FK}:
  \beq\label{a006}
    \displaystyle{
  R^z_{12}(q_{12})
 R^{w}_{23}(q_{23})=R^{w}_{13}(q_{13})R_{12}^{z-w}(q_{12})+
 R^{w-z}_{23}(q_{23})R^z_{13}(q_{13})\,,\quad
 q_{ab}=q_a-q_b\,.
 }
  \eq
Here we use $R$--matrix notations of the quantum inverse scattering method, e.g.
  \beq\label{a007}
    \displaystyle{
  R^z_{12}(q)=\sum\limits_{ijkl=1}^N R_{ij,kl}(z,q)E_{ij}\otimes E_{kl}\otimes 1_N\,,\quad
  R^z_{13}(q)=\sum\limits_{ijkl=1}^N R_{ij,kl}(z,q)E_{ij}\otimes 1_N\otimes E_{kl}\,,
  }
  \eq
 where $E_{ij}$ --- is the standard matrix basis in $\Mat$, $1_N$ ---   is the matrix identity, and $R_{ij,kl}(z,q)$ --- is a set of functions of $z$ and $q$. A normalization of the matrix operator $R^z_{ab}(q_{ab})$  is chosen in a way that for $N=1$ the latter reduces to the scalar function  $\phi(z,q)$ (\ref{a001}).  In this respect equation (\ref{a006}) is a noncommutative generalization of  (\ref{a003}), while the operator $R$~--- is a noncommutative generalization of the Kronecker function.

In addition to(\ref{a006}) one can impose the properties of the skew-symmetry and unitarity
(the latter is a matrix analogue for (\ref{a005})):
  \beq\label{a008}
    \displaystyle{
  R^z_{12}(q)=-R^{-z}_{21}(-q)\,,\qquad
  R^z_{12}(q)R^z_{21}(-q)=1_N\otimes 1_N(\wp(z)-\wp(q))\,.
  }
  \eq
Then such an $R$--operator satisfies the quantum Yang--Baxter equation
  \beq\label{a009}
    \displaystyle{
  R_{12}^\hbar(z_{12})R_{13}^\hbar(z_{13})R_{23}^\hbar(z_{23})=
R_{23}^\hbar(z_{23})R_{13}^\hbar(z_{13})R_{12}^\hbar(z_{12})\,.
 }
  \eq
 Put it differently, a solution of (\ref{a008}) a solution of (\ref{a006})  is a quantum $R$--matrix. Let us mention that even in the scalar case the condition (\ref{a006}) or (\ref{a003})  is very restrictive. At the same time equation (\ref{a009}) is
not restrictive at all since in the scalar case the quantum Yang--Baxter equation is identically true. A class of $R$--matrices satisfying the discussed above conditions includes the elliptic Baxter-Belavin $R$--matrix as well as its trigonometric and rational degenerations, which are equal to the function $\phi(z,q)$ (\ref{a001}) in the scalar case. More detailed description of these $R$--matrices can be found in  \cite{AASZ,LOZ8,Z18}, where an application of this class of $R$--matrices to integrable system is given --- a construction of integrable tops. The main idea goes back to the Sklyanin's paper \cite{Skl1}. He suggested the Hamiltonian description of the classical Euler top by means of the quadratic Poisson algebras, obtained through the classical limit of $RLL$ relations. That is, the classical Euler top was described as the classical limit of $1$ site spin chain. This approach can be developed to obtain explicit description of the Lax pairs with spectral parameters, constructed via the data of $R$--matrices satisfying (\ref{a006}), (\ref{a008}).  A detailed derivation of equations of motion together with the Hamiltonian description by means of the $R$--matrix data is given in papers
  \cite{AASZ} и \cite{LOZ8}  for non-relativistic and relativistic cases respectively.

 \paragraph{Relativistic integrable $GL_N$-top.} In the general case the phase space of $GL_N$-top is given by the set of coordinate functions $S_{ij}$, $i,j=1,\dots,N$ on the Lie group $GL_N$. They are unified into matrix $S=\sum_{ij}S_{ij}E_{ij}$ of size $N\times N$. Equations of motion then take the form the Euler--Arnold equations
 \beq\label{a010}
 \begin{array}{c}
  \displaystyle{
\dot {S}=[{S}, {J}({S})]\,,
 }
 \end{array}
 \eq
where $J(S)$~--- is a linear functional on $S$. It can be written in the form
  \beq\label{a011}
    \displaystyle{
  J(S)=\sum\limits_{i,j,k,l=1}^N J_{ijkl}\,E_{ij}\,S_{lk}\in\Mat
  }
  \eq
  or, using the standard notations $S_1=S\otimes 1_N$, $S_2=1_N\otimes S$,
  \beq\label{a012}
    \displaystyle{
  J(S)=\tr_2(J_{12}S_2)\,,\qquad J_{12}=\sum\limits_{i,j,k,l=1}^N J_{ijkl}\,E_{ij}\otimes E_{kl}\,,
   }
  \eq
 where $\tr_2$ --- is the trace over the second space in the tensor product. Below we give the Lax pair of the relativistic integrable top using the above notation (in the general case, equations (\ref{a010}) are of course not integrable).
For this purpose consider the classical limit of the $R$--matrix:
  \beq\label{a013}
  \begin{array}{c}
      \displaystyle{
R^\hbar_{12}(z)=\frac{1}{\hbar}\,1_N\otimes 1_N+r_{12}(z)+\frac{\hbar}{2}\,\left(r_{12}(z)^2- 1\otimes
1\,\wp(z)\right)+O(\hbar^2) }\,,
  \end{array}
  \eq
 where $r_{12}(z)=-r_{21}(-z)$ --- the classical  $r$--matrix,  and the $\hbar$--order term follows from  (\ref{a008}).
 By comparing this expression with  (\ref{a002}),  we conclude that while the quantum $R$--matrix is a matrix analogue of the Kronecker function, the classical $r$--matrix is a matrix analogue of the first Eisenstein function
$E_1(z)$  (\ref{a001}).
 Consider expansions
  \beq\label{a014}
  \begin{array}{c}
      \displaystyle{
R^{\,z}_{12}(q)=\frac{1}{q}\,P_{12}+R^{z,(0)}_{12}+O(q)\,, \ \
 r_{12}(z)=\frac{1}{z}\,P_{12}+r^{(0)}_{12}+O(z)\,,\ \ P_{12}=\sum\limits_{i,j=1}^N E_{ij}\otimes E_{ji}\,,
}
  \end{array}
  \eq
 where $P_{12}$ --- is the matrix permutation operator.
 Generally speaking, existence of expansions of types (\ref{a013}), (\ref{a014})  is an additional non-trivial requirement for the $R$--matrix. Finally, let us impose one more condition for $R$--matrix:
  \beq\label{a015}
  \begin{array}{c}
      \displaystyle{
R^{\,z}_{12}(q)=R^{\,q}_{12}(z)P_{12}\,.
}
  \end{array}
  \eq
 In the scalar case it turns into equality $\phi(z,q)=\phi(q,z)$. Using (\ref{a015}) and comparing
  (\ref{a013}), (\ref{a014}) we easily get
  \beq\label{a016}
  \begin{array}{c}
      \displaystyle{
 r_{12}(z)=R^{z,(0)}_{12}P_{12}\,.
}
  \end{array}
  \eq
 Now we can formulate the statement on the Lax pair of relativistic top. Namely, for a pair of matrices
  \beq\label{a017}
  \begin{array}{c}
      \displaystyle{
  L(z)=\tr_2(R^\eta_{12}(z)S_2)=\tr_2(R^{\,z}_{12}(\eta)P_{12}S_2)\,,
}
  \end{array}
  \eq
  \beq\label{a018}
  \begin{array}{c}
      \displaystyle{
  M(z)=-\tr_2(r_{12}(z)S_2)=-\tr_2(R^{z,(0)}_{12}P_{12}S_2)
}
  \end{array}
  \eq
the Lax equation
  \beq\label{a000}
  \begin{array}{c}
  \displaystyle{
 \dot{L}(z)=[L(z),M(z)]
 }
 \end{array}
 \eq
is equivalent to equations of motion of the form (\ref{a010}), where
  \beq\label{a019}
  \begin{array}{c}
      \displaystyle{
  J_{12}=R^{\eta,(0)}_{12}-r^{(0)}_{12}\,.
}
  \end{array}
  \eq

\paragraph{Spin generalization of the Ruijsenaars--Schneider model.} In the integrable many-body systems the relativistic generalizations are known as the Ruijsenaars--Schneider models  \cite{Ruijs}.
 We are going to deal with their spin extensions \cite{KrichZ}. The set of dynamical variables consists of particles positions and velocities, and the spin variables are arranged into the matrix $S\in\MatM$. Equations of motion take the following form (for the diagonal and off-diagonal parts of the matrix $S$):
  \beq\label{a020}
  \begin{array}{c}
  \displaystyle{
 {\dot S}_{ii}=-\sum\limits_{k:k\neq i}^M
 S_{ik}S_{ki}\Big( E_1(q_{ik}+\eta)+E_1(q_{ik}-\eta)-2E_1(q_{ik}) \Big)\,,
 }
 \end{array}
 \eq
  \beq\label{a021}
  \begin{array}{c}
  \displaystyle{
 {\dot S}_{ij}=\sum\limits_{k:k\neq j}^M
 S_{ik}S_{kj}\Big( E_1(q_{kj}+\eta)-E_1(q_{kj}) \Big)
 -\sum\limits_{k:k\neq i}^M
 S_{ik}S_{kj}\Big( E_1(q_{ik}+\eta)-E_1(q_{ik}) \Big)
 }
  \end{array}
 \eq
 and
  \beq\label{a0211}
  \begin{array}{c}
  \displaystyle{
 {\ddot q}_i={\dot S}_{ii}\,,
 }
 \end{array}
 \eq
 where $i\neq j$ and $q_{ij}=q_i-q_j$. The Lax pair with spectral parameter
  \beq\label{a022}
  \begin{array}{c}
  \displaystyle{
 L_{ij}(z)=S_{ij}\phi(z,q_{ij}+\eta)\,,\quad i,j=1,...,M;
\quad \res\limits_{z=0}L(z)=S\in\MatM\,,
 }
 \end{array}
 \eq
  \beq\label{a023}
  \begin{array}{c}
  \displaystyle{
 M_{ij}(z)=-\delta_{ij}(E_1(z)+E_1(\eta))S_{ii}-(1-\delta_{ij})S_{ij}\phi(z,q_{ij})\,.
 }
 \end{array}
 \eq
satisfies the Lax equation with additional term (here $\mu_i={\dot q}_i-S_{ii}$)
  \beq\label{a024}
  \begin{array}{c}
  \displaystyle{
 \dot{L}(z)=[L(z),M(z)]+\sum\limits_{i,j=1}^M E_{ij}(\mu_i-\mu_j)S_{ij}f(z,q_{ij}+\eta)\,,\qquad
  f(z,q)=\p_q\phi(z,q)\,,
 }
 \end{array}
 \eq
which turns into zero on-shell constraints
  \beq\label{a025}
  \begin{array}{c}
  \displaystyle{
 \mu_i=0\quad \hbox{or}\quad {S}_{ii}={\dot q}_i\,,\quad i=1,...,M\,.
 }
 \end{array}
 \eq
More precisely, equation (\ref{a024}) is equivalent to (\ref{a020})--(\ref{a021}), and under conditions (\ref{a025}) the Lax equations
with the additional term (\ref{a024})  turn into the ordinary Lax equations (\ref{a000}), and (\ref{a0211}) holds true. Besides the original paper \cite{KrichZ}, the detailed derivation of
(\ref{a024}) can be also found in \cite{Z19}. This derivation is convenient for consideration of a more general system, where the functions entering (\ref{a020})--(\ref{a023}) are replaced by their $R$--matrix analogues.
Although we do not use the Hamiltonian description, let us mention that it is known for the rational and trigonometric systems (see \cite{AF,Resh,Feher,ChF}).

 \paragraph{ The main result of the paper } is the following generalization of simultaneously both --- the relativistic top (\ref{a017})--(\ref{a019}) and the spin Ruijsenaars--Schneider model (\ref{a020})--(\ref{a023}). Consider $\MatNM$--valued Lax pair subdivided into $M\times M$ blocks $ \mL^{ij}(z)=\mL^{ij}(\mS^{ij},z)$ of sizes $N\times N$ each:
 \beq\label{a030}
 \begin{array}{c}
  \displaystyle{
  \mL(z)=\sum\limits_{i,j=1}^M E_{ij}\otimes
  \mL^{ij}(z)\in\MatNM\,,\quad  \mL^{ij}(z)\in\Mat\,,
 }
 \end{array}
 \eq
 \beq\label{a031}
 \begin{array}{c}
  \displaystyle{
  \mL^{ij}(z)=\tr_2 (R_{12}^z(q_{ij} + \eta) P_{12} \mS^{ij}_2)\,,\quad
  \mS^{ij}=\res\limits_{z=0}\mL^{ij}(z)\in\Mat\,,
 }
 \end{array}
 \eq
 \beq\label{a032}
 \begin{array}{c}
  \displaystyle{
  \mM(z)=\sum\limits_{i,j=1}^M E_{ij}\otimes
  \mM^{ij}(z)\in\MatNM\,,\quad  \mM^{ij}(z)\in\Mat\,,
 }
 \end{array}
 \eq
 \beq\label{a033}
 \begin{array}{c}
  \displaystyle{
  \mM^{ij}(z)=-\delta^{ij}\,\tr_2 \Big(R_{12}^{(0), z} P_{12} \mS^{ii}_2\Big)
  -(1-\delta^{ij})\,\tr_2 \Big(R_{12}^z(q_{ij}) P_{12} \mS^{ij}_2\Big)\,.
 }
 \end{array}
 \eq
  The $R$--matrix entering this definition satisfies the associative Yang--Baxter equation (\ref{a006}) together with the properties (\ref{a008}), (\ref{a015}) and the expansions (\ref{a013})--(\ref{a014}).
 Then the Lax equation with the additional term
  \beq\label{a034}
  \begin{array}{c}
  \displaystyle{
 \dot{\mL}(z)=[\mL(z),\mM(z)]+\sum\limits_{i,j=1}^M
 (\mu^i_0-\mu^j_0)\, E_{ij}\otimes\tr_2 \Big(F_{12}^z(q_{ij} + \eta) P_{12} \mS^{ij}_2 \Big)\,,
 }
 \end{array}
 \eq
 where by the analogy with (\ref{a024})
  \beq\label{a0341}
  \begin{array}{c}
  \displaystyle{
F_{12}^z(q)=\p_q R_{12}^z(q)
 }
 \end{array}
 \eq
 and
  \beq\label{a035}
  \begin{array}{c}
  \displaystyle{
 \mu^i_0={\dot q}_i-\tr\Big(\mS^{ii}\Big)\,,\quad i=1,...,M\,,
 }
 \end{array}
 \eq
 is equivalent to equations of motion (in (\ref{a037}) we assume $i\neq j$)
  \beq\label{a036}
  \begin{array}{c}
  \displaystyle{
 {\dot \mS}^{ii}=[\mS^{ii},J^\eta(\mS^{ii})]
 +\sum\limits_{k:k\neq i}^M
 \Big( \mS^{ik}J^{\eta,\,q_{ki}}(S^{ki}) - J^{\eta,\,q_{ik}}(\mS^{ik})S^{ki}
 \Big)\,,
 }
 \end{array}
 \eq
  \beq\label{a037}
  \begin{array}{c}
  \displaystyle{
 {\dot \mS}^{ij}=\mS^{ij}J^\eta(\mS^{jj})-J^\eta(\mS^{ii})\mS^{ij}+
 \sum\limits_{k:k\neq j}^M \mS^{ik}J^{\eta,\,q_{kj}}(\mS^{kj}) -
 \sum\limits_{k:k\neq i}^M J^{\eta,\,q_{ik}}(\mS^{ik})\mS^{kj}\,.
  }
 \end{array}
 \eq
 On-shell constraints
  \beq\label{a038}
  \begin{array}{c}
  \displaystyle{
 \mu^i_0=0\quad \hbox{or}\quad  {\dot q}_i=\tr\Big(\mS^{ii}\Big)\,,\quad i=1,...,M
 }
 \end{array}
 \eq
 (the equations \ref{a034})  reduce to the Lax equations, and
the following equations holds:
  \beq\label{a039}
  \begin{array}{c}
  \displaystyle{
 {\ddot q}_i=\tr\Big({\dot \mS}^{ii}\Big)=\sum\limits_{k:k\neq i}^M
 \tr\Big( \mS^{ik}J^{\eta,\,q_{ki}}(\mS^{ki}) - J^{\eta,\,q_{ik}}(\mS^{ik})\mS^{ki}
 \Big)\,.
 }
 \end{array}
 \eq
 The linear functionals $J^\eta$, $J^{\eta, q}$ from the equations of motion are given by
  \begin{equation}\label{a040}
  \begin{array}{c}
  \displaystyle{
    J^\eta(\mS^{ii}) =  \tr_2 \Big((R_{12}^{(0),\, \eta} - r_{12}^{(0)}) \mS_2^{ii}\Big)\,,
     }
    \\ \ \\
    \displaystyle{
    J^{\eta, q}(\mS^{ij}) =  \tr_2 \Big((R_{12}^{(0),\, q + \eta} - R_{12}^{(0),\, q}) \mS_2^{ij}\Big)\,.
 }
 \end{array}
 \end{equation}
 In the elliptic case the above given Lax pairs and equations of motion are reproduce\footnote{In \cite{Z19}  the elliptic case was described in a slightly different normalization. It differs from the one we use here by $q_j\rightarrow q_j/N$. This
leads to additional factor $1/N$ in the equations of motion in
 \cite{Z19}.} the results of our previous paper  \cite{Z19}, and in the non-relativistic limit the results of
 \cite{GSZ} are reproduced as well. For $N=1$ the $R$--matrix operators under consideration become the scalar functions from (\ref{a001}), thus reproducing the spin Ruijsenaars--Schneider model (\ref{a020})--(\ref{a025}). For $M=1$  the Lax matrices consists of a single block. In this way we come to relativistic top
 (\ref{a017})--(\ref{a019}).
  In the non-relativistic elliptic case the models of the above described type
were first obtained in \cite{Polych},
  They were later described as Hitchin systems on the bundles with non-trivial characteristic classes \cite{LZ}.
  Some explicit examples of the systems can be easily obtained using $R$--matrices used in  \cite{GSZ} in the same normalization as in the present article.


\section{Derivation of equations of motion}\label{sect2}
\setcounter{equation}{0}

\subsection{$R$--matrix identities} To derive equations of motion in the spin Ruijsenaars--Schneider model,
one should use the identity
 (\ref{a004}). Let us rewrite it in a different manner
 \beq\label{a21}
 \begin{array}{c}
  \displaystyle{
   \phi(z,q_1)\phi(z,q_2)=\phi(z,q_1+q_2)(E_1(q_1)+E_1(q_2))-\p_z \phi(z,q_1+q_2)\,,
 }
 \end{array}
 \eq
 where we used that (\ref{a001}) provides $\p_z \phi(z,q)=\phi(z,q)(E_1(z+q)-E_1(z))$. Being written in such a form the identity
 (\ref{a004}) possesses $R$--matrix generalization:
 \beq\label{a22}
 \begin{array}{c}
  \displaystyle{
   R_{12}^z(x) R_{23}^z(y) = R_{13}^z(x + y) r_{12}(x) + r_{23}(y) R_{13}^z(x + y) -
        \frac{\partial}{\partial z} R_{13}^z(x + y)
 }
 \end{array}
 \eq
 Applications of the latter identity can be found in \cite{Z18}. Let us write down its corollary  (\ref{a22}):
  \begin{equation}\label{a222}
  \begin{array}{c}
  \displaystyle{
    R_{12}^z(q_{ik}) R_{23}^z(q_{kj} + \eta) - R_{12}^z(q_{ik} + \eta) R_{23}^z(q_{kj}) =
     }
    \\ \ \\
    \displaystyle{ =
        R_{13}^z(q_{ij} + \eta) (r_{12}(q_{ik}) - r_{12}(q_{ik} + \eta)) +
        (r_{23}(q_{kj} + \eta) - r_{23}(q_{kj})) R_{13}^z(q_{ij} + \eta),
         }
 \end{array}\end{equation}
 We will use degenerations of (\ref{a22}) as well. Consider expansions of its both parts near $x=0$
 \beq\label{a23}
 \begin{array}{c}
  \displaystyle{
   \Big(\frac{1}{x}\, P_{12} + R_{12}^{(0), z} + \ldots\Big) R_{23}^z(y) =
        \Big(R_{13}^z(y) + x F_{13}^z(y) + \ldots\Big) \Big(\frac{1}{x}\, P_{12} + r_{12}^{(0)} + \ldots\Big) + }
        \\ \ \\
        \displaystyle{ +
        r_{23}(y) \Big(R_{13}^z(y) + x F_{13}^z(y) + \ldots\Big) -
        \frac{\partial}{\partial z} \Big(R_{13}^z(y) + x F_{13}^z(y) + \ldots\Big)\,,
 }
 \end{array}
 \eq
where $F_{ab}^z(y)$ is defined as in (\ref{a0341}).  In the zero order in $x$ we have:
 \beq\label{a24}
 \begin{array}{c}
  \displaystyle{
    R_{12}^{(0), z} R_{23}^z(y) = F_{13}^z(y) P_{12} + R_{13}^z(y) r_{12}^{(0)} +
        r_{23}(y) R_{13}^z(y) - \frac{\partial}{\partial z} R_{13}^z(y)\,.
 }
 \end{array}
 \eq
 Also, by expanding  (\ref{a22})  near small $y$ we similarly obtain
 \beq\label{a25}
 \begin{array}{c}
  \displaystyle{
    R_{12}^z(x)
    \Big(\frac{1}{y}\, P_{23} + R_{23}^{(0), z} + \ldots\Big) =
        \Big(R_{13}^z(x) + y F_{13}^z(x) + \ldots\Big) r_{12}(x) + }\\ \ \\ \displaystyle{ +
        \Big(\frac{1}{y}\, P_{23} + r_{23}^{(0)} + \ldots\Big) \Big(R_{13}^z(x) + y F_{13}^z(x) + \ldots\Big) -
        \frac{\partial}{\partial z}  \Big(R_{13}^z(x) + y F_{13}^z(x) + \ldots\Big)\,,
 }
 \end{array}
 \eq
 \beq\label{a26}
 \begin{array}{c}
  \displaystyle{
    R_{12}^z(x) R_{23}^{(0), z} = R_{13}^z(x) r_{12}(x) + r_{23}^{(0)} R_{13}^z(x) +
        P_{23} F_{13}^z(x) - \frac{\partial}{\partial z} R_{13}^z(x)\,.
 }
 \end{array}
 \eq
 From (\ref{a24}) and (\ref{a26}) we deduce
  \begin{equation}\label{a27}
  \begin{array}{c}
  \displaystyle{
    R_{12}^{(0), z} R_{23}^z(q_{ij} + \eta) - R_{12}^z(\eta) R_{23}^z(q_{ij}) =
     }
    \\ \ \\
    \displaystyle{ =
    F_{13}^z(q_{ij} + \eta) P_{12} + R_{13}^z(q_{ij} + \eta) (r_{12}^{(0)} - r_{12}(\eta)) +
    (r_{23}(q_{ij} + \eta) - r_{23}(q_{ij})) R_{13}^z(q_{ij} + \eta)\,.
     }
 \end{array}
 \end{equation}

\subsection{Lax equation}
Let us write down the Lax equation with the  additional term (\ref{a034})
explicitly in terms of $N\times N$ blocks. For the diagonal blocks this yields
 \beq\label{a30}
 \begin{array}{c}
  \displaystyle{
  \dot{\mL}_{ii}(z) =
        \mL^{ii}(z) \mM^{ii}(z) - \mM^{ii}(z) \mL^{ii}(z) +
        \sum_{k \neq i} \Big(\mL^{ik}(z) \mM^{ki}(z) - \mM^{ik}(z) \mL^{ki}(z)\Big)\,.
 }
 \end{array}
 \eq
Similarly, for the off-diagonal part we have
 \beq\label{a31}
 \begin{array}{c}
  \displaystyle{
  \dot{\mL}_{ij}(z) =
        \mL^{ii}(z) \mM^{ij}(z) - \mM^{ii}(z) \mL^{ij}(z) +
        \mL^{ij}(z) \mM^{jj}(z) - \mM^{ij}(z) \mL^{jj}(z) + }\\ \ \\ \displaystyle{ +
        \sum_{k \neq i, j} \Big(\mL^{ik}(z) \mM^{kj}(z) - \mM^{ik}(z) \mL^{kj}(z)\Big)
        +(\mu^i_0-\mu^j_0)\,\tr_2 \Big(F_{12}^z(q_{ij} + \eta) P_{12}  \mS^{ij}_2 \Big)\,.
 }
 \end{array}
 \eq
 Our aim is to show that (\ref{a30}) and (\ref{a31}) are equivalent to the equations of motion
 (\ref{a036}) and (\ref{a037}) respectively.
  Notice that
 \beq\label{a32}
 \begin{array}{c}
  \displaystyle{
 \res\limits_{z=0}\mL(z)= S=-\res\limits_{z=0}\mM(z)\in\MatNM\,,
 }
 \end{array}
 \eq
i.e. the second order pole in $z$ is cancelled out in commutator $[\mL(z),\mM(z)]$.

\paragraph{Off-diagonal part.} In the l.h.s. of (\ref{a31}) we have
 \begin{equation}\label{a40}
 \begin{array}{c}
 \displaystyle{
    \dot{\mL}^{ij}(z)_1 =  \tr_2 (F_{12}^z(q_{ij} + \eta) P_{12}  \mS^{ij}_2 \dot{q}_{ij}) +
                         \tr_2 (R_{12}^z(q_{ij} + \eta) P_{12} \dot{\mS}^{ij}_2)\,.
 }\end{array}
 \end{equation}
 The index 1 in the l.h.s. means that the Lax equation is in the first tensor component. Consider expression in the r.h.s. of (\ref{a31})
 \begin{equation}\label{a41}
 \begin{array}{c}
 \displaystyle{
    (\mL^{ik}(z) \mM^{kj}(z) - \mM^{ik}(z) \mL^{kj}(z))_1 = }\\ \ \\ \displaystyle{ =
     \tr_{23} \Big(-R_{12}^z(q_{ik} + \eta) P_{12} \mS_2^{ik} R_{13}^z(q_{kj}) P_{13} \mS_3^{kj} +
              R_{12}^z(q_{ik}) P_{12} \mS_2^{ik} R_{13}^z(q_{kj} + \eta) P_{13} \mS_3^{kj}\Big) = }\\ \ \\ \displaystyle{ =
     \tr_{23} \Big(\Big(R_{12}^z(q_{ik}) R_{23}^z(q_{kj} + \eta) - R_{12}^z(q_{ik} + \eta) R_{23}^z(q_{kj})\Big)
              P_{12} \mS_2^{ik} P_{13} \mS_3^{kj}\Big)=
 }
 \\ \ \\
 \displaystyle{
    \stackrel{(\ref{a222})}{=}
     \tr_{23} \Big(R_{13}^z(q_{ij} + \eta) \Big(r_{12}(q_{ik}) - r_{12}(q_{ik} + \eta)\Big)
              P_{12} \mS_2^{ik} P_{13} \mS_3^{kj}\Big) + }\\ \ \\ \displaystyle{ +
     \tr_{23} \Big(\Big(r_{23}(q_{kj} + \eta) - r_{23}(q_{kj})\Big) R_{13}^z(q_{ij} + \eta)
              P_{12} \mS_2^{ik} P_{13} \mS_3^{kj}\Big)\,.
 }\end{array}
 \end{equation}
 Two obtained terms are transformed using (\ref{a015})--(\ref{a016}) and the permutation operator property
 $P_{12}U_{12}= U_{21}P_{12}$ (and  $P_{12}U_{23}= U_{13}P_{12}$ respectively). Let us transform the first term from the r.h.s. of  (\ref{a41}):
 \begin{equation}\label{a42}
 \begin{array}{c}
 \displaystyle{
     \tr_{23} \Big(R_{13}^z(q_{ij} + \eta) \Big(r_{12}(q_{ik}) - r_{12}(q_{ik} + \eta)\Big)
              P_{12} \mS_2^{ik} P_{13} \mS_3^{kj}\Big) = }\\ \ \\ \displaystyle{ =
    -  \tr_3 \Big(R_{13}^z(q_{ij} + \eta) P_{13} P_{13}\,
         \tr_2 \Big((R_{12}^{(0), q_{ik} + \eta} - R_{12}^{(0), q_{ik}}) \mS_2^{ik}\Big)
        P_{13} \mS_3^{kj}\Big) = }\\ \ \\ \displaystyle{ =
    -  \tr_3 \Big(R_{13}^z(q_{ij} + \eta) P_{13}
         \,\tr_2 \Big((R_{32}^{(0), q_{ik} + \eta} - R_{32}^{(0), q_{ik}}) \mS_2^{ik}\Big)
        \mS_3^{kj}\Big) = }\\ \ \\ \displaystyle{ =
    -  \tr_2 \Big(R_{12}^z(q_{ij} + \eta) P_{12}
         \,\tr_3 \Big((R_{23}^{(0), q_{ik} + \eta} - R_{23}^{(0), q_{ik}}) \mS_3^{ik}\Big)
        \mS_2^{kj}\Big).
 }\end{array}
 \end{equation}
 Using the defintion (\ref{a040}) we obtain
 \begin{equation}\label{a43}
 \begin{array}{c}
 \displaystyle{
    \tr_{23} \Big(R_{13}^z(q_{ij} + \eta) \Big(r_{12}(q_{ik}) - r_{12}(q_{ik} + \eta)\Big)
              P_{12} \mS_2^{ik} P_{13} \mS_3^{kj}\Big) =
                }
              \\ \ \\
              \displaystyle{ =
    -  \tr_2 \Big(R_{12}^z(q_{ij} + \eta) P_{12} {J^{\eta, q_{ik}}( \mS^{ik})}_2\, \mS_2^{kj}\Big)\,.
 }
 \end{array}
 \end{equation}
 The second term from the r.h.s. of (\ref{a41}) is transformed in a similar manner:
  $$
 \begin{array}{c}
 \displaystyle{
     \tr_{23}  \Big( \Big(r_{23}(q_{kj} + \eta) - r_{23}(q_{kj}) \Big) R_{13}^z(q_{ij} + \eta)
              P_{12} \mS_2^{ik} P_{13} \mS_3^{kj} \Big) = }\\ \ \\ \displaystyle{ =
     \tr_{23}   \Big(R_{13}^z(q_{ij} + \eta) P_{12} \mS_2^{ik} P_{13} \mS_3^{kj}
               \Big(r_{23}(q_{kj} + \eta) - r_{23}(q_{kj} \Big) \Big) = }\\ \ \\ \displaystyle{ =
     \tr_{23}  \Big(R_{13}^z(q_{ij} + \eta) P_{13} P_{23} \mS_2^{ik} \mS_3^{kj}
         \Big(R_{23}^{(0), q_{kj} + \eta} - R_{23}^{(0), q_{kj}} \Big) P_{23} \Big) =
       }\end{array}
 $$
  $$
 \begin{array}{c}
 \displaystyle{
          =
     \tr_{23}  \Big(P_{23} R_{13}^z(q_{ij} + \eta) P_{13} P_{23} \mS_2^{ik} \mS_3^{kj}
         \Big(R_{23}^{(0), q_{kj} + \eta} - R_{23}^{(0), q_{kj}} \Big) \Big) = }\\ \ \\ \displaystyle{ =
     \tr_{23}  \Big(R_{12}^z(q_{ij} + \eta) P_{12} \mS_2^{ik} \mS_3^{kj}
         \Big(R_{23}^{(0), q_{kj} + \eta} - R_{23}^{(0), q_{kj}} \Big) \Big) =
    }\end{array}
 $$
 \begin{equation}\label{a44}
 \begin{array}{c}
  \displaystyle{
          =
     \tr_2  \Big(R_{12}^z(q_{ij} + \eta) P_{12} \mS_2^{ik}
         \,\tr_3\,  \Big(\mS_3^{kj} (R_{23}^{(0), q_{kj} + \eta} - R_{23}^{(0), q_{kj}}) \Big) \Big) = }
         \\ \ \\ \displaystyle{ =
     \tr_2  \Big(R_{12}^z(q_{ij} + \eta) P_{12} \mS_2^{ik} J^{\eta, q_{kj}}( \mS^{kj})_2 \Big)\,.
 }\end{array}
 \end{equation}
  Finally, from (\ref{a43}) and (\ref{a44})  we get the following answer for initial expression (\ref{a41}):
 \begin{equation}\label{a45}
 \begin{array}{c}
 \displaystyle{
    (\mL^{ik}(z) \mM^{kj}(z) - \mM^{ik}(z) \mL^{kj}(z))_1 =
     }
    \\ \ \\
     \displaystyle{ =
     \tr_2  \Big(R_{12}^z(q_{ij} + \eta) P_{12}
         \Big( \mS^{ik} J^{\eta, q_{kj}}( \mS^{kj}) - J^{\eta, q_{ik}}( \mS^{ik})  \mS^{kj} \Big)_2 \Big)\,.
 }
 \end{array}\end{equation}
  Next, consider the following expression from  (\ref{a31}):
 \begin{equation}\label{a46}
 \begin{array}{c}
 \displaystyle{
    (\mL^{ii}(z) \mM^{ij}(z) - \mM^{ii}(z) \mL^{ij}(z))_1 = }\\ \ \\ \displaystyle{ =
     \tr_{23} \Big(-R_{12}^z(\eta) P_{12} \mS_2^{ii} R_{13}^z(q_{ij}) P_{13} \mS_3^{ij} +
              R_{12}^{(0), z} P_{12} \mS_2^{ii} R_{13}^z(q_{ij} + \eta) P_{13} \mS_3^{ij}\Big) = }\\ \ \\ \displaystyle{ =
     \tr_{23} \Big(\Big(R_{12}^{(0), z} R_{23}^z(q_{ij} + \eta) - R_{12}^z(\eta) R_{23}^z(q_{ij})\Big)
              P_{12} \mS_2^{ii} P_{13} \mS_3^{ij}\Big)\,.
 }
 \end{array}
 \end{equation}
 Apply relation (\ref{a27}):
 \begin{equation}\label{a47}
 \begin{array}{c}
 \displaystyle{
    (\mL^{ii}(z) \mM^{ij}(z) - \mM^{ii}(z) \mL^{ij}(z))_1 =
     \tr_{23} \Big(F_{13}^z(q_{ij} + \eta) P_{12} P_{12} \mS_2^{ii} P_{13} \mS_3^{ij}\Big) + }\\ \ \\ \displaystyle{ +
     \tr_{23} \Big(R_{13}^z(q_{ij} + \eta)
       \Big (r_{12}^{(0)} - r_{12}(\eta)\Big) P_{12} \mS_2^{ii} P_{13} \mS_3^{ij}\Big) + }\\ \ \\ \displaystyle{ +
     \tr_{23} \Big(\Big(r_{23}(q_{ij} + \eta) - r_{23}(q_{ij})\Big) R_{13}^z(q_{ij} + \eta)
        P_{12} \mS_2^{ii} P_{13} \mS_3^{ij}\Big)\,.
 }\end{array}
 \end{equation}
 Let us simplify all three terms in the r.h.s. of (\ref{a47}). Transform the first term:
\begin{equation}\label{a48}
\begin{array}{c}\displaystyle{
     \tr_{23} \Big(F_{13}^z(q_{ij} + \eta) P_{12} P_{12} \mS_2^{ii} P_{13} \mS_3^{ij}\Big) =
     \tr_2 \Big(\mS_2^{ii}\Big)\,  \tr_3 \Big(F_{13}^z(q_{ij} + \eta) P_{13} \mS_3^{ij}\Big) =
    }
    \\ \ \\
    \displaystyle{
    =
     \tr  \mS^{ii} \cdot  \tr_2 \Big(F_{12}^z(q_{ij} + \eta) P_{12} \mS_2^{ij}\Big)\,.
    }
    \end{array}
\end{equation}
 The third term is already known:
 \begin{equation}\label{a49}
 \begin{array}{c}
 \displaystyle{
     \tr_{23} \Big(\Big(r_{23}(q_{ij} + \eta) - r_{23}(q_{ij})\Big) R_{13}^z(q_{ij} + \eta)
        P_{12} \mS_2^{ii} P_{13} \mS_3^{ij}\Big) =
        }
        \\ \ \\
        \displaystyle{
     =\tr_2 \Big(R_{12}^z(q_{ij} + \eta) P_{12} \mS_2^{ii} J^{\eta, q_{ij}}( \mS^{ij})_2\Big)\,.
     }
      \end{array}
 \end{equation}
 For the second term from the r.h.s. of  (\ref{a47}) we obtain
 \begin{equation}\label{a50}
 \begin{array}{c}
 \displaystyle{
     \tr_{23} \Big(R_{13}^z(q_{ij} + \eta)
        \Big(r_{12}^{(0)} - r_{12}(\eta)\Big) P_{12} \mS_2^{ii} P_{13} \mS_3^{ij}\Big) = }\\ \ \\ \displaystyle{ =
     \tr_{23} \Big(R_{13}^z(q_{ij} + \eta) P_{13}
        \Big(r_{32}^{(0)} - r_{32}(\eta)\Big) P_{32} \mS_2^{ii} \mS_3^{ij}\Big) = }\\ \ \\ \displaystyle{ =
     \tr_2 \Big(R_{12}^z(q_{ij} + \eta) P_{12}
         \,\tr_3 \Big\{\Big(r_{23}^{(0)} - r_{23}(\eta)\Big) P_{23} \mS_3^{ii}\Big\} \mS_2^{ij}\Big) = }\\ \ \\ \displaystyle{ =
        -  \tr_2 \Big(R_{12}^z(q_{ij} + \eta) P_{12} J^\eta( \mS^{ii})_2 \mS_2^{ij}\Big)\,.
 }
 \end{array}
 \end{equation}
 Thus, the expression (\ref{a47})  takes the form:
 \begin{equation}\label{a51}
 \begin{array}{c}
 \displaystyle{
    (\mL^{ii}(z) \mM^{ij}(z) - \mM^{ii}(z) \mL^{ij}(z))_1 = }\\ \ \\ \displaystyle{ =
     \tr_2 \Big(R_{12}^z(q_{ij} + \eta) P_{12}
        \Big( \mS^{ii} J^{\eta, q_{ij}}( \mS^{ij}) - J^\eta( \mS^{ii}\Big)  \mS^{ij})_2\Big) +
        }\\ \ \\ \displaystyle{+
     \tr  \mS^{ii} \cdot  \tr_2 \Big(F_{12}^z(q_{ij} + \eta) P_{12} \mS_2^{ij}\Big)\,.
 }
 \end{array}
 \end{equation}
 One more expression $\mL^{ij}(z) \mM^{jj}(z) - \mM^{ij}(z) \mL^{jj}(z)$ from (\ref{a31}) is transformed similarly to (\ref{a47}). This yields
 \begin{equation}\label{a52}
 \begin{array}{c}
 \displaystyle{
    (\mL^{ij}(z) \mM^{jj}(z) - \mM^{ij}(z) \mL^{jj}(z))_1 = }\\ \ \\ \displaystyle{ =
     \tr_2 \Big(R_{12}^z(q_{ij} + \eta) P_{12}
        \Big( \mS^{ij} J^\eta( \mS^{jj}) - J^{\eta, q_{ij}}( \mS^{ij})  \mS^{jj}\Big)_2\Big) -
        }
        \\ \ \\ \displaystyle{-
     \tr  \mS^{jj} \cdot  \tr_2 \Big(F_{12}^z(q_{ij} + \eta) P_{12} \mS_2^{ij}\Big)\,.
 }
 \end{array}
 \end{equation}
 Collecting the terms (\ref{a45}), (\ref{a51}) and (\ref{a52}) gives the following answer for
 $ij$--block of the commutator:
 \begin{equation}\label{a53}
 \begin{array}{c}
 \displaystyle{
    ([\mL(z), \mM(z)]^{ij})_1 =
    ( \tr  \mS^{ii} -  \tr  \mS^{jj}) \cdot  \tr_2 (F_{12}^z(q_{ij} + \eta) P_{12} \mS_2^{ij}) +
     \tr_2 (R_{12}^z(q_{ij} + \eta) P_{12} A_2), }\\ \ \\ \displaystyle{
    A =  \mS^{ii} J^{\eta, q_{ij}}( \mS^{ij}) - J^\eta( \mS^{ii})  \mS^{ij} +
         \mS^{ij} J^\eta( \mS^{jj}) - J^{\eta, q_{ij}}( \mS^{ij})  \mS^{jj} +
        }\\ \ \\ \displaystyle{
        +
        \sum_{k \ne i, j} \Big( \mS^{ik} J^{\eta, q_{kj}}( \mS^{kj}) - J^{\eta, q_{ik}}( \mS^{ik})  \mS^{kj}\Big)\,.
 }
 \end{array}
 \end{equation}
 Also, taking into account the last term (with $\mu^0_i$) in the r.h.s. of  (\ref{a31}),  we get (\ref{a021}) in the form
 \begin{equation}\label{a54}
 \begin{array}{c}
 \displaystyle{
    \dot{\mS}^{ij} =  \mS^{ii} J^{\eta, q_{ij}}( \mS^{ij}) - J^\eta( \mS^{ii})  \mS^{ij} +
         \mS^{ij} J^\eta( \mS^{jj}) - J^{\eta, q_{ij}}( \mS^{ij})  \mS^{jj} +\phantom{\Big(}
         }\\  \displaystyle{
        +\sum_{k \ne i, j}^M \Big( \mS^{ik} J^{\eta, q_{kj}}( \mS^{kj}) - J^{\eta, q_{ik}}( \mS^{ik})  \mS^{kj}\Big)\,.
 }
 \end{array}
 \end{equation}

Let us comment on transition from (\ref{a53}) to (\ref{a54}). Strictly speaking, we have proved that the Lax hold true on equations of motion but we have not proved the inverse. In order to prove the inverse statement we need to see that all components of the matrix equation
 (\ref{a54})  are contained in (\ref{a53}) independently taking also into account that $R_{12}$ is a linear operator, which may mix these components somehow. Put it differently, we need to show that
$\tr_2(R_{12}^z(q_{ij}+\eta)P_{12}C_2)=0$ leads to $C=0$.
For this purpose consider the Lax equation near $z=0$. It follows from (\ref{a013})--(\ref{a015}) that $R_{12}^z(q_{ij}+\eta)P_{12}$ has a simple pole in $z=0$ and the residue is equal to $P_{12}$. Then the desired statement follows from $\tr_2(P_{12}A_2)=A$.


 \paragraph{Diagonal part.} Consider now the equation (\ref{a30}), which l.h.s. is of the form:
 \begin{equation}\label{a64}
 \begin{array}{c}
 \displaystyle{
    \dot{\mL}_{ii}(z)_1 =  \tr_2 (R_{12}^z(\eta) P_{12} \dot{\mS}^{ii}_2)\,.
 }\end{array}
 \end{equation}
 In the r.h.s. we transform the expression under sum using (\ref{a222}):
 \begin{equation}\label{a65}
 \begin{array}{c}
 \displaystyle{
    (\mL^{ik}(z) \mM^{ki}(z) - \mM^{ik}(z) \mL^{ki}(z))_1 = }\\ \ \\ \displaystyle{ =
     \tr_{23} \Big(-R_{12}^z(q_{ik} + \eta) P_{12} \mS_2^{ik} R_{13}^z(q_{ki}) P_{13} \mS_3^{ki} +
              R_{12}^z(q_{ik}) P_{12} \mS_2^{ik} R_{13}^z(q_{ki} + \eta) P_{13} \mS_3^{ki}\Big) =
              }\\ \ \\ \displaystyle{ =
     \tr_{23} \Big(\Big(R_{12}^z(q_{ik}) R_{23}(q_{ki} + \eta) - R_{12}^z(q_{ik} + \eta) R_{23}^z(q_{ki})\Big)
        P_{12} \mS_2^{ik} P_{13} \mS_3^{ki}) =
         }
   \end{array}
 \end{equation}
   $$
    \begin{array}{c}
        \displaystyle{ \stackrel{(\ref{a222})}{=}
     \tr_{23} \Big(R_{13}^z(\eta) \Big(r_{12}(q_{ik}) - r_{12}(q_{ik} + \eta)\Big)
        P_{12} \mS_2^{ik} P_{13} \mS_3^{ki}\Big) + }\\ \ \\ \displaystyle{ +
     \tr_{23} \Big(\Big(r_{23}(q_{ki} + \eta) - r_{23}(q_{ki})\Big) R_{13}^z(\eta)
        P_{12} \mS_2^{ik} P_{13} \mS_3^{ki}\Big) = }\\ \ \\ \displaystyle{ =
     \tr_2 \Big(R_{12}^z(\eta) P_{12}
        ( \mS^{ik} J^{\eta, q_{ki}}( \mS^{ki}) - J^{\eta, q_{ik}}( \mS^{ik})  \mS^{ki})_2\Big)\,.
 }
 \end{array}
 $$
 The rest of the expression in the r.h.s. of (\ref{a30}) is simplified via (\ref{a27}):
 \begin{equation}\label{a75}
 \begin{array}{c}
 \displaystyle{
    (\mL^{ii}(z) \mM^{ii}(z) - \mM^{ii}(z) \mL^{ii}(z))_1 = }\\ \ \\ \displaystyle{ =
     \tr_{23} \Big(-R_{12}^z(\eta) P_{12} \mS_2^{ii} R_{13}^{(0), z} P_{13} \mS_3^{ii} +
              R_{12}^{(0), z} P_{12} \mS_2^{ii} R_{13}^z(\eta) P_{13} \mS_3^{ii}\Big) = }\\ \ \\ \displaystyle{ =
     \tr_{23} \Big(\Big(R_{12}^{(0), z} R_{23}^z(\eta) - R_{12}^z(\eta) R_{23}^{(0), z}\Big)
        P_{12} \mS_2^{ii} P_{13} \mS_3^{ii}\Big) = }\\ \ \\ \displaystyle{ \stackrel{(\ref{a27})}{=}
     \tr_{23} \Big(R_{13}^z(\eta) \Big(r_{12}^{(0)} - r_{12}(\eta)\Big)
        P_{12} \mS_2^{ii} P_{13} \mS_3^{ii}\Big) + }\\ \ \\ \displaystyle{ +
     \tr_{23} \Big(\Big(r_{23}(\eta) - r_{23}^{(0)}\Big) R_{13}^z(\eta)
        P_{12} \mS_2^{ii} P_{13} \mS_3^{ii}\Big) + }\\ \ \\ \displaystyle{ +
     \tr_{23} \Big(F_{13}^z(\eta) P_{12} P_{12} \mS_2^{ii} P_{13} \mS_3^{ii}\Big) -
     \tr_{23} \Big(P_{23} F_{13}^z(\eta) P_{12} \mS_2^{ii} P_{13} \mS_3^{ii}\Big)\,.
 }
 \end{array}
 \end{equation}
 Notice that the two last terms are equal, so that they are cancelled out:
 \begin{equation}\label{a76}
 \begin{array}{c}
 \displaystyle{
     \tr_{23} \Big(F_{13}^z(\eta) P_{12} P_{12} \mS_2^{ii} P_{13} \mS_3^{ii}\Big) =
     \tr_{23} \Big(P_{23} F_{13}^z(\eta) P_{12} \mS_2^{ii} P_{13} \mS_3^{ii}\Big) = }\\ \ \\ \displaystyle{ =
     \tr  \mS^{ii} \cdot  \tr_2 \Big(F_{12}^z(\eta) P_{12} \mS_2^{ii}\Big)\,.
 }
 \end{array}
 \end{equation}
 The first and the second terms from  (\ref{a75}) are of the form
 \begin{equation}\label{a85}
 \begin{array}{c}
 \displaystyle{
     \tr_{23} \Big(R_{13}^z(\eta) \Big(r_{12}^{(0)} - r_{12}(\eta)\Big)
        P_{12} \mS_2^{ii} P_{13} \mS_3^{ii}\Big) + }\\ \ \\ \displaystyle{ +
     \tr_{23} \Big(\Big(r_{23}(\eta) - r_{23}^{(0)}\Big) R_{13}^z(\eta)
        P_{12} \mS_2^{ii} P_{13} \mS_3^{ii}\Big) =
        }\\ \ \\ \displaystyle{ =
     \tr_2 \Big(R_{12}^z(\eta) P_{12} \Big(J^\eta( \mS^{ii})  \mS^{ii} -  \mS^{ii} J^\eta( \mS^{ii})\Big)_2\Big)\,.
 }
 \end{array}
 \end{equation}
 Finally, from (\ref{a65}) and (\ref{a85}) we get the answer
 \begin{equation}\label{a86}
 \begin{array}{c}
 \displaystyle{
    ([L(z), M(z)]_{ii})_1 =  \tr_2 \Big(R_{12}^z(\eta) P_{12} B_2\Big)\,,
 }\end{array}
 \end{equation}
 where
 \begin{equation}\label{a861}
 \begin{array}{c}
     \displaystyle{
    B = J^\eta( \mS^{ii})  \mS^{ii} -  \mS^{ii} J^\eta( \mS^{ii}) +
        \sum_{k \ne i} \Big( \mS^{ik} J^{\eta, q_{ki}}( \mS^{ki}) - J^{\eta, q_{ik}}( \mS^{ik})  \mS^{ki}\Big)\,.
 }\end{array}
 \end{equation}
 Here one should also use the argument given after (\ref{a54}).
 Thus, the equations of motion are verified for the diagonal blocks.

\subsection{Interacting tops}
As was explained in \cite{Z19}, in the particular case $\mathrm{rk}(\mS)=1$ the equations of motion
can be written in terms of the diagonal blocks only.
Let us recall the main idea. The property $\mathrm{rk}(\mS)=1$ yields
 \beq\label{a87}
 \begin{array}{c}
  \displaystyle{
   \mS^{ik}_1 P_{12} \mS^{ki}_1=\mS^{ii}_1 \mS^{kk}_2\,.
  }
 \end{array}
 \eq
 Next, for an arbitrary $J(S)=\tr_2(J_{12}S_2)$  of the form (\ref{a012})
 and ${\breve J}_{12}=J_{12}P_{12}$ we have
 \beq\label{a88}
 \begin{array}{c}
  \displaystyle{
  J(S)=\tr_2(J_{12}S_2)=\tr_2({\breve
  J}_{12}P_{12}S_2)
  =\tr_2(S_2{\breve J}_{12}P_{12})=
 }
 \\ \ \\
  \displaystyle{
 =\tr_2(S_2P_{12}{\breve J}_{21})=\tr_2(P_{12}S_1{\breve
 J}_{21})\,.
  }
 \end{array}
 \eq
Therefore
 \beq\label{a89}
 \begin{array}{c}
  \displaystyle{
  \mS^{ik}J(\mS^{ki})
  =\tr_2\Big( \mS^{ik}_1 P_{12} \mS^{ki}_1{\breve J}_{21}
  \Big)=\mS^{ii} \tr_2\Big( {\breve
  J}_{21}\,\mS^{kk}_2 \Big)\,,
  }
 \end{array}
 \eq
where
 \beq\label{a90}
 \begin{array}{c}
  \displaystyle{
   {\breve  J}_{21}=P_{12}{\breve  J}_{21}P_{12}=P_{12}J_{12}\,.
  }
 \end{array}
 \eq
In the same way
 \beq\label{a91}
 \begin{array}{c}
  \displaystyle{
  J(\mS^{ik})\mS^{ki}=
  \tr_2\Big( {\breve J}_{12} \mS^{ik}_1 P_{12} \mS^{ki}_1 \Big)
  = \tr_2\Big( {\breve
  J}_{12}\,\mS^{kk}_2 \Big)\mS^{ii}\,.
 }
 \end{array}
 \eq
Finally, equations (\ref{a036}) and (\ref{a039}) are written in the form
  \beq\label{a92}
  \begin{array}{c}
  \displaystyle{
 {\dot \mS}^{ii}=[\mS^{ii},J^\eta(\mS^{ii})]
 +\sum\limits_{k:k\neq i}^M
 \Big( \mS^{ii}{\widetilde J}^{\eta,\,q_{ki}}(S^{kk}) - {\breve J}^{\eta,\,q_{ik}}(\mS^{kk})S^{ii}
 \Big)\,,
 }
 \end{array}
 \eq
  \beq\label{a93}
  \begin{array}{c}
  \displaystyle{
 {\ddot q}_i=\tr\Big({\dot \mS}^{ii}\Big)=
 \sum\limits_{k:k\neq i}^M
 \tr\Big( \mS^{ii}{\widetilde J}^{\eta,\,q_{ki}}(S^{kk}) - {\breve J}^{\eta,\,q_{ik}}(\mS^{kk})S^{ii}
 \Big)\,,
 }
 \end{array}
 \eq
where
  \beq\label{a94}
 \begin{array}{c}
  \displaystyle{
 {\widetilde J}^{\eta,\,q_{ki}}(\mS^{kk})=
 \tr_2\Big({\breve J}_{21}^{\eta,\,q_{ki}}\mS^{kk}_2 \Big)=\tr_2\Big(P_{12}{ J}_{12}^{\eta,\,q_{ki}}\mS^{kk}_2 \Big)\,,
 }
 \end{array}
 \eq
  \beq\label{a95}
 \begin{array}{c}
  \displaystyle{
 {\breve J}^{\eta,\,q_{ik}}(\mS^{kk})=
  \tr_2\Big({\breve J}_{12}^{\eta,\,q_{ik}}\mS^{kk}_2 \Big)=\tr_2\Big({J}_{12}^{\eta,\,q_{ik}}P_{12}\,\mS^{kk}_2 \Big)\,.
 }
 \end{array}
 \eq
 Being written in the form (\ref{a92})--(\ref{a93}) the equations of motion can be treated as dynamics of $M$ particles
bearing additional ''spin'' type degrees of freedom, i.e. the particles can be identified with the tops, which have also positions and velocities besides their own internal degrees of freedom. The interaction between the tops depends on the distance and on the spin dynamical variables.


\subsubsection*{Acknowledgments}
This research (including the results in Sec. 2) was performed at the Steklov Mathematical Institute of Russian
Academy of Sciences and is supported by a grant from the Russian Science Foundation (Project No. 19-11-00062).

\begin{small}

\end{small}

\end{document}